\documentclass[12pt]{iopart}

\usepackage{graphicx}
\usepackage{color}
\usepackage{bm}

\begin{document}

\title[]{Dynamics of ultrafast heated radiative plasmas driven by petawatt laser lights}

\author{K. Sugimoto$^{a,b}$, N. Iwata$^{b,c}$, A. Sunahara$^{d}$, T. Sano$^{b}$ and Y. Sentoku$^{b}$}

\address{
$^{a}$Department of Physics, Graduate School of Science, Osaka University, 1-1 Machikanecho, Toyonaka, Osaka 560-0043, Japan\\
$^{b}$Institute of Laser Engineering, Osaka University, 2-6 Yamadaoka, Suita, Osaka 565-0871, Japan\\
$^c$Institute for Advanced Co-Creation Studies, Osaka University, 1-1 Yamadaoka, Suita, Osaka 565-0871, Japan\\
$^{d}$CMUXE, School of Nuclear Engineering, Purdue University, Indiana 47907, USA
}%

\ead{sugimoto-k@osaka-u.ac.jp}
\vspace{10pt}

\begin{abstract}
A relativistic petawatt laser light can heat heavy metals over keV temperature isochorically and ionize them almost fully. 
Copious hard X-rays are emitted from the high-Z hot plasma which acts as X-ray sources, while they work as a cooling process of the plasma. 
The cooling process can affect on the creation of high energy density plasma via the interaction, however, the details are unknown.
The X-ray spectrum depends on the plasma temperature, so that it is worthwhile to investigate the radiation cooling effects.
We here study the isochoric heating of a solid silver foil irradiated by relativistic laser lights with a help of particle-in-cell simulations including Coulomb collisions, ionizations, and radiation processes. 
We have conducted a parameter survey varying laser intensity, $10^{18-20}\,\rm{W/cm^2}$, to check the cooling effects while keeping the incident laser energy constant.
The silver plasma heated mainly by the resistive heating dissipates its energy by keV X-ray emissions in a picosecond time scale. 
The radiation power from the silver foil is found to be comparable to the incident laser power when the laser intensity is less than $10^{19}\,{\rm W/cm^2}$ under the constant energy situation. 
The evolution of the plasma energy density inside the target is then suppressed, due to which a highly compressed collisional shock is formed at the target surface and propagates into the plasma. 
The radiation spectra of the keV silver plasma are also demonstrated. 
\end{abstract}

\section{\label{sec:level1}Introduction}
The advent of petawatt laser systems enables us to explore the high field science and the high energy density science in a wide parameter range. 
Kilo-joule (kJ) class petawatt lasers, e.g. LFEX and NIF-ARC, having picoseconds pulse durations heat dense plasmas over peta-pascal pressures \cite{MatsuoPRL2020}, and also accelerate particles more efficiently than short pulse laser lights with the similar intensity \cite{Yogo2017,Mariscal2019}.
The other type of petawatt laser lights has extremely short pulse duration of a few tens femtoseconds and a micron-revel focal spot with an intensity over $10^{21}$\,W/cm$^2$. 
The laser plasma interaction with such a ultrashort and extreme laser light, e.g. J-KAREN-P and ELI-beamlines, initiates the QED processes accompanying copious gamma-rays and subsequent pair productions. 
Recently the laser-driven heavy ion acceleration via the target normal sheath acceleration (TNSA) had been demonstrated using a thin silver foil irradiated by J-KAREN-P laser light \cite{Nishiuchi2020}.
The silver ions are ionized over charge states 40 and they are accelerated by the TNSA at the target rear surface.
The ionization degree of the heavy ions is an important factor for the efficient acceleration.
There are some numerical works to evaluate the ionization degrees of heavy metals in the laser-ion acceleration \cite{Nishiuchi2020,Kawahito2020,Domanski2020} and in the isochoric heating \cite{Mishra2013}. 

Since such intense laser and high-$Z$ plasma interactions accompany a considerable amount of hard X-rays emissions, it is critically important to understand the characteristics of the radiations and their effect on the plasma heating and the laser-ion acceleration.
Understanding the radiation properties of the hot heavy metal is a key to realize high brightness X-ray sources for various applications including a diagnostic of the dense plasma core in the laser fusion.
Although Mishra {\it et al.}\,\cite{Mishra2013} had considered the radiation effects in the isochoric heating, the model considered only the integrated radiation powers in the photon energies.
To clarify the radiation effects on the laser-high $Z$ plasma interaction including the radiation properties is the main motivation of the study.

\begin{figure}[t!]
 \centering
 \includegraphics[width=8cm]{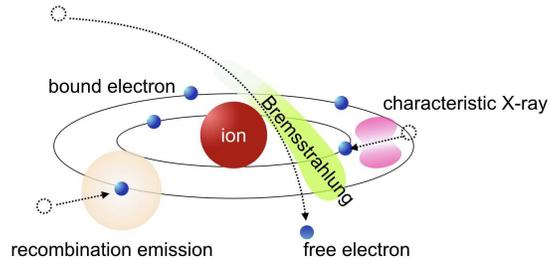}
 \caption{Radiation processes considered in this paper, Bremsstrahlung, recombination emission and characteristic X-ray.}
 \label{rad_pro}
\end{figure}  

Hard X-rays with photon energies over keV are emitted in heavy metal targets when the petawatt laser light is irradiated on the target surface. 
Energetic electrons injected into the target drive the bulk plasma heating and also the radiation processes illustrated in Fig.\,\ref{rad_pro}.
When the electron collides with an atom or partially ionized ion, a bound electron might be kicked out, and the outer bound electron subsequently comes down to the inner electron orbit by emitting a X-ray photon (characteristic X-ray).
The characteristic X-ray between K- and L-shells is called as $K_\alpha$. 
When the hot electron suffers a large angle scattering by the nucleus, the electron changes the direction and emits photons with continuous spectrum (bremsstrahlung).
The photon energies via bremsstrahlung depends on the atomic number $Z$, as $Z^{2}$ and the $K_{\alpha}$ photon energy increases as $Z$ increases \cite{Hou2006}.
When the free electron is recombined to the orbit, a X-ray photon is also emitted with energy depending on both the bound orbit and the energy of the free electron before recombination. 

The time integrated conversion efficiency to hard X-rays from the input laser energy have been studied experimentally \cite{Park2006,Borm2019}, and the temporal scales of the X-ray emissions are observed to be around the laser pulse durations. 
However, the time evolution of the radiative plasmas cannot be investigated from the observations.
The energy loss due to radiation decreases the electron temperature, which will affect the creation of the radiative plasmas.

This paper is organized in the following way. 
Section II presents the radiation model coupled to the particle-in-cell (PIC) simulations and also the theoretical estimation of the isochoric heating and the radiation cooling effect. 
In Section III we present one-dimensional (1D) and two-dimensional (2D) PIC simulation results and compare the simulation results to the theoretical estimation.
Section IV is dedicated to the discussion and summary.

\section{\label{sec:level2}Radiation model in PIC simulations}

We use a PIC code, PICLS, which features binary collisions among charged particles, dynamic ionization processes in gas and solid density plasmas, and also radiation processes \cite{Mishra2013,SentokuJCP08,Pandit2012,Sentoku2014}.
In the present work, we included bremsstrahlung, radiative recombination, and radiative de-excitation illustrated in Fig.\,\ref{rad_pro} in a thin silver plasma.
The emissivities of these three radiative processes were computed by the Collisional Radiative Equilibrium (CRE) model \cite{Saltzmann} for various ion densities and electron temperatures. 
We computed the emissivities for plasmas with ion densities from $10^{20}$ to $10^{24}$\,cm$^{-3}$ and electron temperatures from 1\,eV to 100\,keV.
The photon energy range considered in the calculations is from 1\,eV to 100\,keV.
In the simulations, we calculate the radiated energy by linearly interpolating in log$_{10}$ scale the database with the average electron temperature and ion density in each PIC cell.
The emitted energy is subtracted from the bulk electron energy in the cell, so that the plasma is cooled down by the radiation emission. 

\begin{figure}[t!]
 \centering
 \includegraphics[width=8cm]{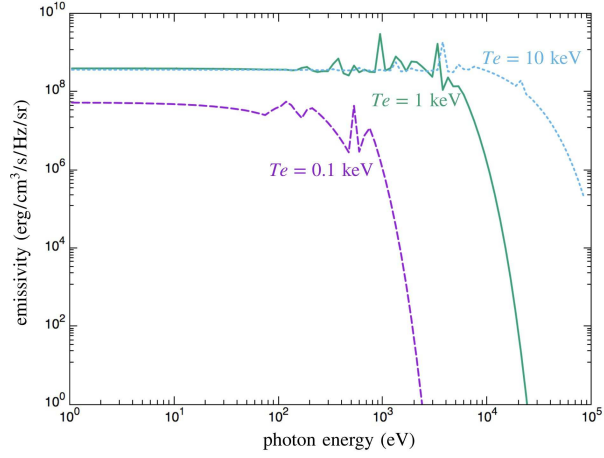}
 \caption{Emissivity spectra of solid silver plasma at electron temperatures of 0.1, 1, and 10\,keV. Some peaks represent the characteristic X-rays.}
 \label{cre_spect}
\end{figure}

Figure\,\ref{cre_spect} shows the emissivity spectra from a solid silver plasma at electron temperatures of $T_e = $ 0.1, 1, and 10\,keV.
Most part of the spectra consist of continuous components which correspond to bremsstrahlung and radiative recombination processes.
Spikes correspond to the characteristic X-rays.
For higher $T_e$, the high energy continuous component via bremsstrahlung and recombination increases, and consequently, the fraction of the characteristic X-rays decreases.

We have studied a silver plasma heating by a short pulse laser light using the 1D PICLS simulation including the radiation cooling \cite{Sugimoto2020}. 
We here extend the analysis to study the radiation effect on the plasma dynamics including the multidimensional effect by conducting 2D PICLS simulations.
In this study, we consider only an optically thin plasma. 
The temperature of the target reaches above keV via the interaction.
In a solid plasma at such high temperature, the length with optical thickness equal to 1\,${\rm \mu m}$ for hard X-rays is a several microns.

Before going to the simulations, here, we estimate the impact of the radiation cooling during the isochoric heating of a thin silver foil using the emissivities from the CRE model.
When the thin foil is irradiated by a relativistic petawatt laser light, the hot electrons are generated at the interaction surface and they flow into the target with mega-ampere current.
The hot electron current is then neutralized by the bulk return current, and thus the target is heated mainly by the resistive (Joule) heating over keV temperature.
The achievable electron temperature $T_e\,(\rm  keV)$ via the resistive heating is derived as \cite{Leblanc2014},
\begin{eqnarray}
\centering
T_e(a_0,\bar{t}) \,\simeq\, 511\times\left[\frac{5}{3}\frac{\eta_{0}L\left(1+\chi{a_0}^2/2\right)\bar{t}}{\bar{n}_i}\right]^{2/5},
\label{resi_heat}
\end{eqnarray}
as a function of the normalized laser amplitude $a_0=eE_L/(m_e c \omega_0)$ and the interaction time $\bar{t}$ normalized by the laser oscillation period $2\pi/\omega_0$. 
Here, $\eta_{0}$ is a constant $e^{2}\omega_{0}/m_{e}c^{3} \simeq 1.6\times10^{-8}$ where $e$ the elementary charge, $m_e$ the electron rest mass, and $c$ the speed of light. $L$ is the Coulomb logarithm, $\chi$ is the absorption coefficient, $\bar{n}_i$ is the ion density normalized by the critical density $n_c=\omega_0^2 m_e/4\pi e^2$, and $E_L$ is the laser electric field amplitude.

\begin{figure}[b!]
 \centering
 \includegraphics[width=8cm]{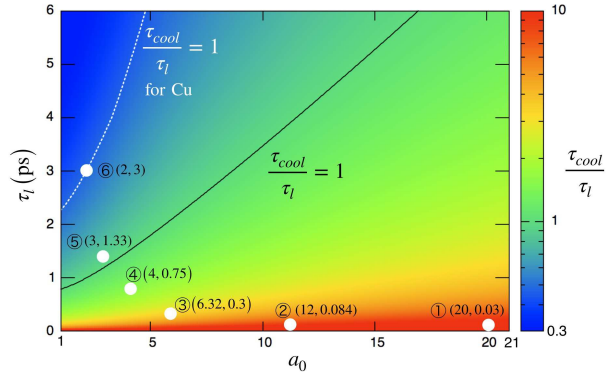}
  \caption{
Theoretical estimation of the radiation cooling effect on laser heating of solid silver target. The color indicates the ratio of the cooling time $\tau_{cool}$ (Eq.\,\ref{rad_cooling}) to the laser pulse duration $\tau_l$. The black solid and white dashed lines show $\tau_{cool}/\tau_l=1$ for silver and copper (Cu) targets, respectively.
} 
  \label{map}
\end{figure} 

We can estimate the radiation power density $P$ from the CRE database using $T_e$ given by Eq.\,\ref{resi_heat} at the time when the laser irradiates the target over time $\bar{t}=\bar{\tau}_l$ where $\bar{\tau}_{l}$ is the laser pulse duration normalized by the laser oscillation period.
The absorbed energy is stored in the plasma as the electron energy density $n_eT_e$. 
We define the cooling time $\tau_{\rm cool}$ as the time scale the plasma energy is taken away by the radiation as
\begin{eqnarray}
 {\tau}_{\rm cool}\equiv\frac{n_{e}T_{e}(a_0,\bar{\tau}_l)}{P\left(n_{i},T_{e}(a_0,\bar{\tau}_l)\right)}.
 \label{rad_cooling}
\end{eqnarray}
If the laser pulse duration ${\tau}_{l}$ is less than ${\tau}_{\rm cool}$, the radiation cooling is negligible during the laser heating.
While, if ${\tau}_{l}$ is longer than ${\tau}_{\rm cool}$, the radiation cooling affects seriously on the temporal evolution of the bulk electron temperature.

We plotted the ratio $\tau_{\rm cool}/\tau_{l}$ as a function of $a_0$ and $\tau_l$ in Fig.\,\ref{map}.
The reddish area indicates the laser heating power exceeds the radiation power, while the blueish area means the radiation power overcomes the laser heating power.
The solid line represents the laser condition where the both powers becomes equal, i.e. $\tau_{\rm cool}/\tau_{l}=1$. 
In the region above this line, the laser isochoric heating will be degraded by the radiation cooling. 
Note that the region dominated by the radiation cooling ($\tau_{\rm cool}/\tau_{l}<1$) is seen for the over-picosecond pulse durations $\tau_{l}>1$\,ps for the silver target.
We also plot a white dashed line which corresponds to the border $\tau_{\rm cool}/\tau_{l}=1$ for a copper (Cu) target. 
It can be seen that radiations from a Cu foil are negligible for most of the current high intensity laser lights.
Heavier metals, e.g. gold, have the radiation cooling effect with a shorter pulse duration. 

\begin{figure}[t!]
 \centering
 \includegraphics[width=8cm]{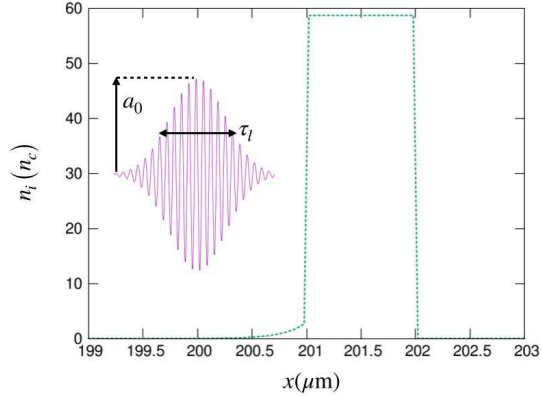}
  \caption{Simulation setup configuration for 1D PIC simulations. The incident laser pulse comes from the left boundary. The silver foil target is composed of pre-plasma ($x=200-201\,{\rm \mu m}$) and solid region ($x= 201-202\,{\rm \mu m}$).}
  \label{setup}
\end{figure}

In the following 1D-PICLS simulations, we chose 6 laser profiles as shown in Fig.\,\ref{map} by the white circles. 
For all the cases, we keep the laser energy constant, but change the laser amplitude and pulse duration.
These laser parameters are achievable by the modern laser facilities. 
Figure\,\ref{setup} shows the simulation setup configuration. 
The 1D simulations have the system size $402\,{\rm \mu m}$ and a 1\,$\mu$m silver foil target with ion density $n_{i}=58n_{c}$ at the center. 
The target is composed of pre-plasma ($x=200-201\,{\rm \mu m}$) with a scale length of 0.2\,{$\rm \mu m$}, and solid foil ($x=201-202\,{\rm \mu m}$).
The initial charge state of the silver target is $Z_i=1$. 
The ionization potential of the silver ion is set by the NIST database \cite{NIST}.
The incident laser wavelength is fixed to {$1\, \rm \mu m$}.
We have varied $a_0$ from 20 to 2 and pulse duration $\tau_l$ from $30\,{\rm fs}$ to  $3\,{\rm ps}$ with the Gaussian pulse profile with maintaining the input laser energy constant. 
The peak amplitude reaches to the foil surface $x=201\,{\rm \mu m}$ at $t=0$.
We include the Coulomb collisions, field ionizations, and impact ionizations in all the simulations. 
While the radiation calculations are on and off to see the impact of the radiation loss.
The multi-dimensional effects are checked with the 2D-PICLS code.
In the 2D simulations, we used the same laser pulse profile as in 1D calculation with $a_{0}=2, \tau_{l}=3.0\,{\rm ps}$.
The simulation box size is $42\times20\,{\rm \mu m^{2}}$, and the laser light is focused on the center of the target with a spot diameter $3\,{\rm \mu m}$.

\section{\label{sec:level3}Simulation Results}

\subsection{1D: radiation cooling}

\begin{figure}[t!]
 \centering
 \includegraphics[width=8cm]{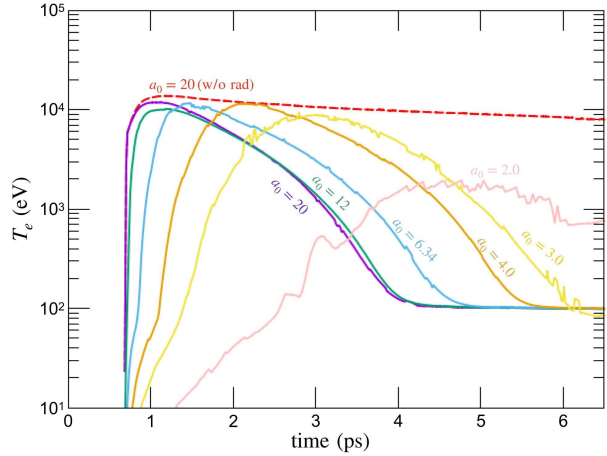}
  \caption{The temporal evolution of electron temperature in the 1D simulations. The temperature was measured in the region where ion density was over 29\,$n_c$ which is the half of solid density. The solid lines and dashed line are for the simulations with and without the radiation cooling, respectively.}
  \label{temp_evol}
\end{figure}

We conducted 1D simulations to investigate the radiation cooling effect.
Figure\,\ref{temp_evol} shows the temporal evolution of the bulk electron temperature for the simulations with the radiation cooling.
The temperature was measured in the region where the ion density was over 29\,$n_{c}$ which corresponds to the half of the initial solid density.
We see the degradation of the peak bulk temperature $T_e$ for the lower intensity cases.  
For the case with $a_0=2.0$ and $\tau_l=3$\,ps, $T_e$ saturates around keV even though the laser pulse is still irradiated. 
It is also seen that the life time of 10\,keV silver plasma is a few picoseconds. 
This infers that the plasma internal energy will be converted to the hard X-rays in a few picosecond time scale.
We also show the result for $a_0=20$ and $\tau_l=30$\,fs without the radiation loss by the red dashed line. 
In this case, the temperature decreases only via the adiabatic cooling due to the plasma expansion. 
We see that the adiabatic cooling is much slower than the radiative cooling. 
The conversion rate from the laser energy to the hard X-ray emissions, which is discussed later, is about $1\,\%$ for the current laser configurations. 
When the electron temperature drops to $100\,{\rm eV}$, the emissivity decreases significantly as shown in Fig.\,\ref{cre_spect}.
Hence, at the time later than $t=4\,{\rm ps}$, the temperature stays at about $100\,{\rm eV}$ in the cases of $a_{0}=3-20$. 

In the present research, we have included only non-relativistic bremsstrahlung emissions.
Therefore, hot electrons with energies over $100\,{\rm keV}$ do not suffer from the radiation energy loss.
Relativistic electrons emit photons with energies over MeV, so called $\gamma$-rays via bremsstrahlung or the radiation damping.
The energy conversion rate for $\gamma$-rays is much less than $1\,\%$ \cite{Pandit2012}, so that $\gamma$-rays do not affect the plasma heating. 

\begin{figure}[t!]
 \centering
 \includegraphics[width=7.5cm]{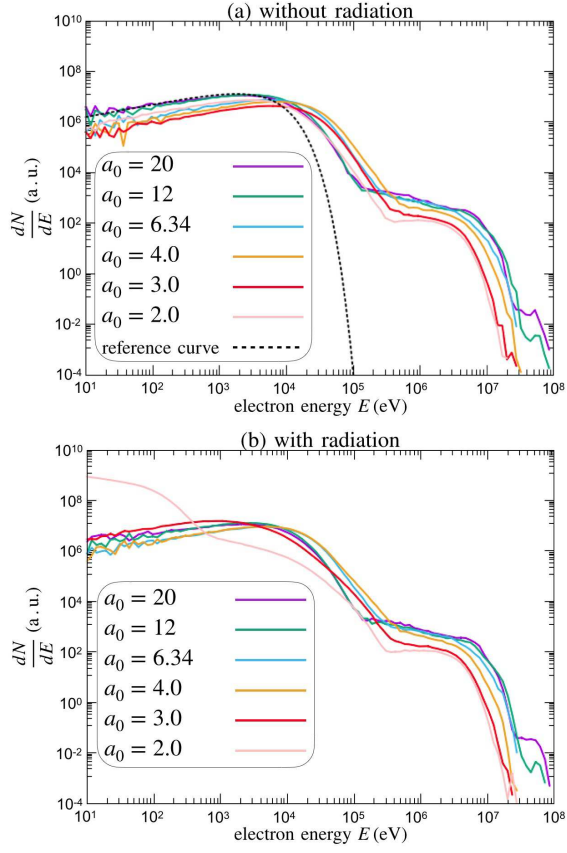}
  \caption{Electron energy spectra in the 1D simulations (a) without and (b) with the radiation cooling. The spectrum is observed when the laser pulse irradiation ends in each case.}
  \label{e_spect}
\end{figure}

Figures\,\ref{e_spect} (a) and (b) show electron energy spectra calculated at the moment when the laser irradiation was over for the cases without radiations (a) and with radiations (b).
The spectra have two parts, one for the bulk electrons with energies $E < 100\,{\rm keV}$ and the other for hot electrons with energies $E > 100\,{\rm keV}$.
The peak energy of the bulk electrons is explained by the resistive heating. 
The Maxwellian distribution $\propto E^{1/2}\,{\rm exp}\left(-E/T_{e}\right)$ with the bulk electron temperature $T_e$ calculated from Eq.\,(\ref{resi_heat}) for $a_0=20$ and $\tau_l=30$\,fs is plotted by the dotted line as a reference. 
We see that the reference line agrees well with the bulk electron component of $a_0=20$. 
Note here that from Eq.\,(\ref{resi_heat}), we see the $T_e$ slightly increases with the weaker amplitude and longer pulse duration but it is almost same around 10\,keV for the same pulse energy cases.
The hot electrons are induced in the laser-plasma interaction at the target surface. 
Their average energy is given by the ponderomotive scaling as
$\varepsilon_{p}=[ (1+a_{0}^{2}/2)^{1/2} -1 ]m_{e}c^2$.
For example, $a_{0}=20$ derives $\varepsilon_{p}\sim6.7\,{\rm MeV}$ \cite{Wilks1992}. 
The hot electron temperature decreases as $a_{0}$ decreases as seen in the spectra in the energy range $E > 100\,{\rm keV}$. 

The radiation cooling effect on the electron spectra is seen in Fig.\,\ref{e_spect} (b).
Compared with Fig.\,\ref{e_spect} (a), the peak bulk electron energy of $a_{0}=$ 3.0 and 2.0 shift to the lower energy, while other cases do not, as predicted in the discussion of Fig.\,\ref{map}.

\subsection{1D: effect of radiation cooling on plasma dynamics}

\begin{figure}[b!]
 \centering
 \includegraphics[width=8.0cm]{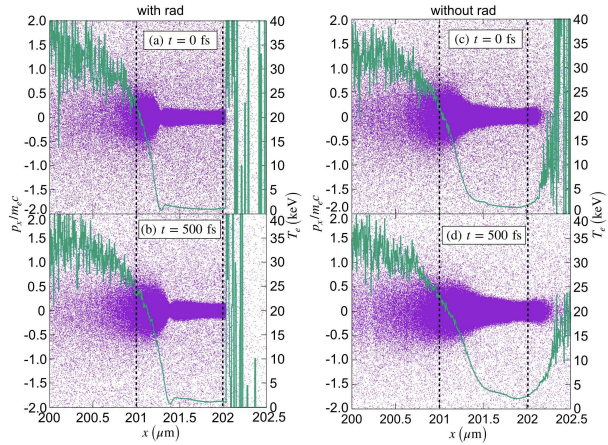}
  \caption{Electron energy temperature distribution (solid lines) and electron phase (dots) with (a), (b) and without (c), (d) radiations for the laser light with $a_0=2$ and $\tau_l=3$\,ps. The dot lines show the initial foil surface positions.}
  \label{e_temp}
\end{figure}

Since the plasma dynamics reflect how the radiations cool down the plasma, it is worthwhile to investigate the dynamics under a situation in which the radiations take the plasma energy significantly.
Figure\,\ref{e_temp} shows the temporal evolution of electron temperature $T_e$ (green solid lines) and electron phase $x$-$p_x$ plot (dots) around the target observed at the moments when the laser peak reached the solid surface ($t=0\,{\rm fs}$) and at $t=500$\,fs.
The laser is the long pulse with $a_0=2$ and $\tau_l=3$\,ps.
The plots (a) and (b) represent the results with radiations, and (c) and (d) represent those without radiations.
Initially, the silver foil is located in the region $x=201 - 202\,{\rm \mu m}$.

In the interaction region with the laser light ($x<201$\,$\mu$m), $T_e$ is similar not depending on whether the radiations are taken into account or not. 
This is because the hot plasma expanding in front of the dense plasma region has a low density, so that the dominant radiation via bremsstrahlung, whose power is proportional to the square of the ion density, is negligible.
In the solid region ($x>201.5$\,$\mu$m), $T_e$ in the case with the radiations stays low around 100\,eV while $T_e$ without the radiations is hotter exceeding 1\,keV.
The temperature has a gradient due to the thermal diffusion from the hot region. The electron momentum distribution inside the solid density region shrinks in the vertical direction in (a) and (b) due to the radiation. 

\begin{figure}[t!]
 \centering
 \includegraphics[width=8.5cm]{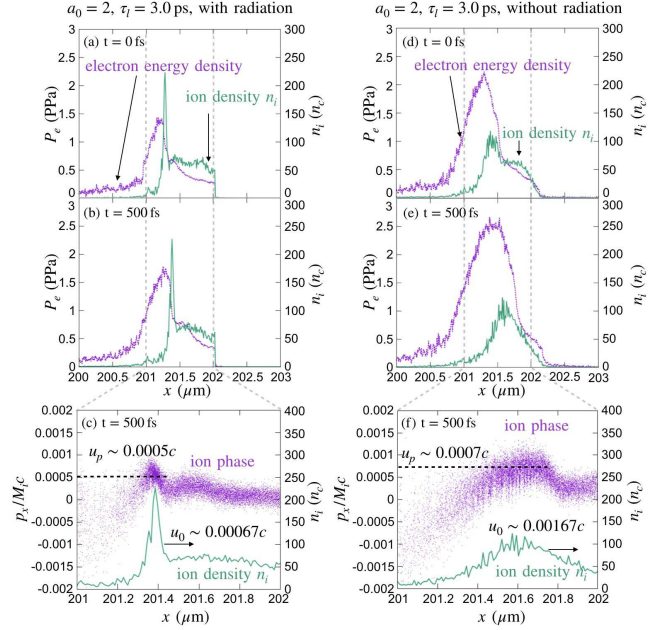}
  \caption{Temporal evolution of ion density (green) and electron energy density (purple) with (a), (b) and without (d), (e) radiations. Ion phase (dots) and ion density (lines) are shown in (c) and (f) for the cases with and without radiations, respectively. The incident laser parameters are $a_0=2$ and $\tau_l=3$\,ps.}
  \label{iondens_edens}
\end{figure}

Figures\,\ref{iondens_edens} (a), (b), (d), and (e) show the temporal evolution of ion densities $n_i$ (green lines) normalized by $n_c$ and electron energy densities (PPa) (purple lines) observed at the same time in the same simulations of Fig.\,7.
The broken lines indicate the initial foil region. (a)-(c) are the plots for the simulations with radiations, and (d)-(f) are those for the simulations without radiations. 

At $t=0$, the foil surface is pushed forward by the hot electrons' pressure. 
The compressed ion density $n_i$ at the foil surface in (a) is about 4 times higher than the initial density. 
The increase of $n_i$ is significantly higher than the case without radiations.
Due to the radiation cooling, the electron pressure inside the solid region decreases. Hence, the bulk plasma can be compressed more easily with the radiation cooling effect. 
A strong radiation happens in the compressed region, which results a dent in the temperature profile at $x=201.4\,{\rm \mu m}$ in Fig.\,\ref{e_temp} (b).

Figures\,\ref{iondens_edens} (c) and (f) display ion phase $x-p_{x}$ diagram and ion density distribution at $t=500\,{\rm fs}$.
The ion momentum is normalized by $M_{i}c$ where $M_{i}$ is the ion mass.
It can be seen that the collisional shock is formed because of the pressure gradient. 
Since the upstream pressures in both with and without radiations are similar, 
the speed of the compressed surface $u_p$ is similar, $u_p\sim 0.0005c$ with radiations and $0.0007c$ without radiations as indicated by the dotted lines in (c) and (f). 
The acoustic speed at $T_e =$1\,keV is $C_s=(T_e Z_i/M_i)^{1/2}\simeq 0.00055c$ with $Z_{i}=30$ for the case with radiations and $C_s\simeq 0.00063$ with $Z_i=40$ for the case without radiations where $Z_i$ is the average ion charge state in each simulation.
The shock speed $u_0$ is explained well by $u_0=u_p\rho_1/(\rho_1-\rho_0)$ \cite{Zeldovich} where $\rho_{0}$ and $\rho_{1}$ are uncompressed and compressed ion densities, respectively, through the shock surface.
In the case with radiations, $u_0\sim 0.00067c$ which is 3 times slower than the case without radiations.
And thus the target with radiation takes 3 times longer time to expand into vacuum than the case without radiation.
We also confirmed the momentum and energy conservations through the shock surface \cite{Zeldovich}. 
These results show that the radiation cooling affects the plasma dynamics by reducing the shock speed in the dense plasma.

\subsection{1D: hard X-ray emissions}\label{sec-xray}

\begin{figure}[t!]
 \centering
 \includegraphics[width=8cm]{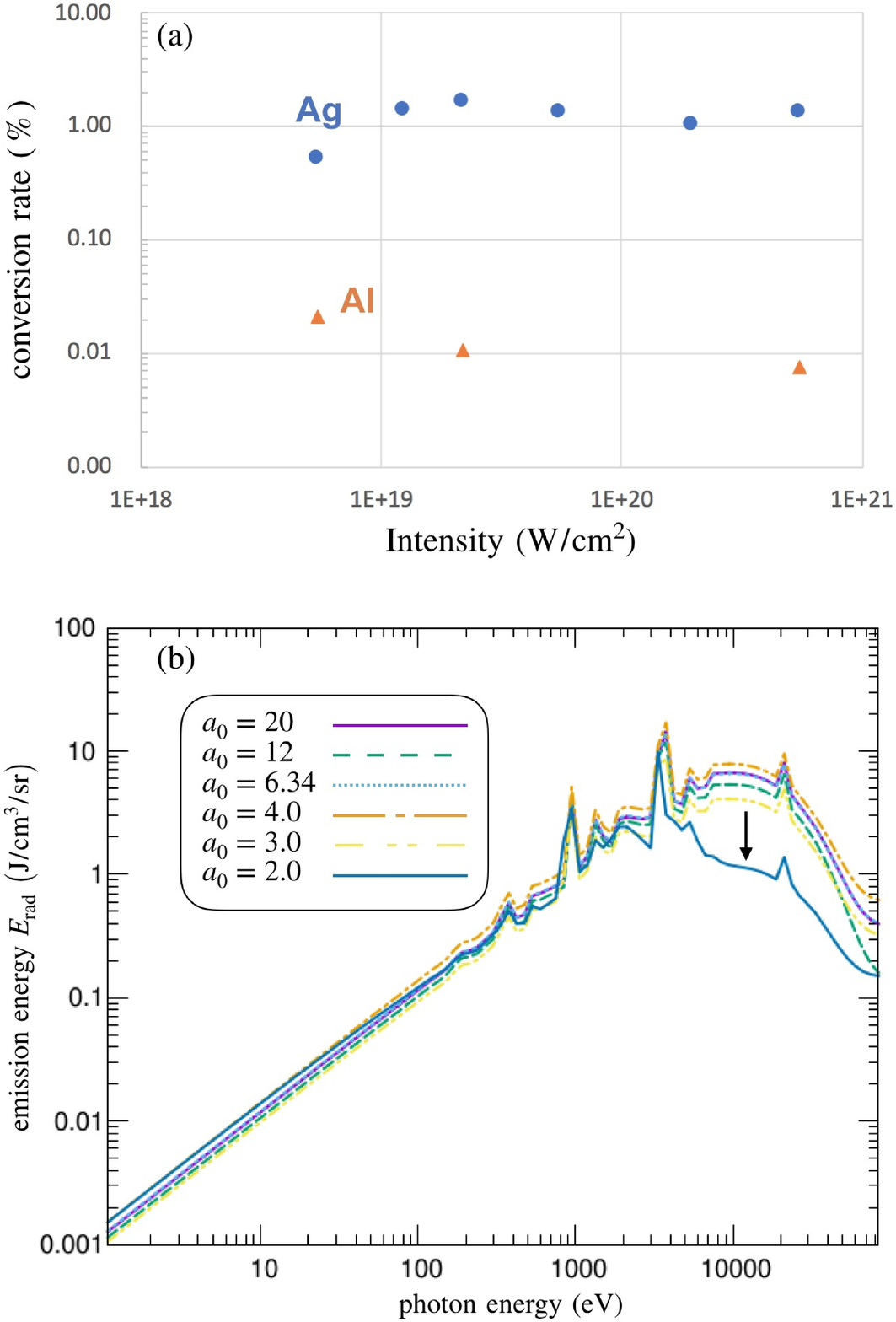}
  \caption{(a) Conversion efficiency from the laser light to hard X-rays with photon energy of $18-75\,{\rm keV}$ for the Ag (circles) and Al (triangles) foil targets. (b) Spectra of the total emission energy for each simulations. The emission energy is integrated spatially over the simulation box and temporally over the total simulation time.}
  \label{xray_conv}
\end{figure}

A hot metal plasma with keV electron temperature has a potential to be a hard X-ray source.
We calculated the conversion efficiency from the laser light to hard X-ray photons with energies of 18-75\,${\rm keV}$.
The photons with this energy range could be a backlight source for the high density plasma experiments \cite{Borm2019}.
Figure\,\ref{xray_conv} (a) shows the conversion efficiency of Ag (circles) and Al (triangles) foils with thickness of 1\,$\mu$m.
The emitted X-ray energy was obtained by integrating the X-ray emissivity over photon frequency and solid angle in each PIC cell and summing the energy during the simulations.
It is found that the conversion efficiency of Ag is nearly two-orders of magnitude higher than that of Al.
The bulk electron temperatures achieved by the laser irradiation are almost same around 10\,keV in both materials.
At such high temperature, the Al plasma is fully ionized and the Ag plasma is almost fully ionized to the charge state of 45.

Below we explain the high X-ray yields in the Ag target. 
The bremsstrahlung is the dominant radiation process and its emission power density $E_{ff}$\,(W/cm$^3$) depends on the charge state $Z_i$ of the plasma as
$E_{ff}=1.69\times10^{-32}n_{e}T_{e}^{1/2}\sum_{Z_{i}}\left(Z_{i}^{2}n_i(Z_{i})\right)$
where $n_{e}$ is electron number density (1/cm$^3$), $T_{e}$ is electron temperature (eV), and $n_i(Z_{i})$ is ion number density (1/cm$^3$) with charge state $Z_{i}$ \cite{2016nrl}.
If we assume the charge neutral state, $n_{e}=\sum_{Z_{i}}\left(Z_{i}n_{i}(Z_{i})\right)$.
Since the ratio of the atomic number of Ag and Al is $47/13$, the radiation power of bremsstrahlung from the Ag plasma is $\sim (47/13)^{3} = 47.11$ times higher than that from the Al plasma.
In addition to bremsstrahlung, strong emissions through the radiative recombination process are expected in the Ag plasma. 
The power of the radiative recombination emission of Ag is close to that of bremsstrahlung, so that the Ag plasma has about two-orders of magnitude higher emission than the Al plasma. 

Note here that the dependence of the conversion efficiency on the laser intensity in Fig.,\ref{xray_conv} (a) is weak in the intensity regime $>10^{19}\,{\rm W/cm^2}$, since the achievable electron temperature via the resistive heating is not much changed when the total laser energy is same. 
In the case of laser intensity below $10^{19}$ W/cm$^{2}$, the efficiency decreases to a half of other cases.
This is due to the radiation cooling that occurs in the picosecond time scale during the laser plasma interaction. 
Namely, the electron temperature cannot reaches keV-level in the dense plasma region with radiations as seen in Figs.\,6 and 7, and thus over-keV X-ray emissions are suppressed.
Figure\,\ref{xray_conv} (b) shows emission energy spectra in the simulations with the different laser amplitudes.
Here, the spectra for photon energy $h\nu_{i}$ were calculated as
\begin{eqnarray}
 {E_{\rm rad}(h\nu_{\rm i}) }\ =\int_{ T_{\rm sim}}\sum_{\rm j}( \eta_{\rm i}\times h\Delta \nu_{\rm i})\,dt 
\label{spect_cal}
\end{eqnarray}
where $\rm i$ and $\rm j$ denote indices of the photon energy group and the spatial grid, respectively.
$\eta_{\rm i}$ is emissivity of photon energy of $h\nu_{\rm i}$ where $h$ and $\nu$ are Planck constant and photon frequency, respectively, and $\Delta\nu_{\rm i}=\nu_{\rm i}-\nu_{\rm i-1}$.
The time integration is taken over the total simulation time $T_{\rm sim}$.
Figure\,\ref{xray_conv} (b) indicates that the photons with energies above $1\,{\rm keV}$ from bremsstrahlung and recombination processes mainly contribute to the cooling.

We see that the emission energy over 3\,keV decreases significantly in the case of $a_0=2$ due to the radiation cooling as indicated with an arrow in Fig.\,\ref{xray_conv} (b).
This result is consistent with the lower conversion efficiency seen in Fig.\,\ref{xray_conv} (a). 
The emissivity of bremsstrahlung is proportional to 
$T_{e}^{-1/2}\exp(-h\nu/T_{e})$
 \cite{1979radpro}, and therefore, the peak of the emission energy appears around the energy comparable to the electron temperature.

\subsection{2D simulation}

\begin{figure}[b!]
 \centering
 \includegraphics[width=8cm]{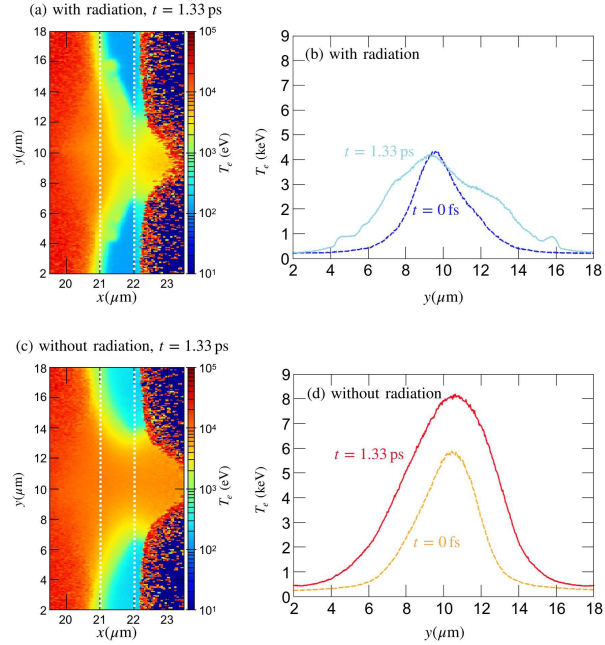}
  \caption{Electron temperature distribution in the 2D PIC simulations with (a), (b) and without (c), (d) radiations. (a) and (c) show the temperature at the moment the laser peak arrived at the foil surface. The white dashed lines present the initial foil surfaces. (b) and (d) show the temporal evolution of the electron temperature averaged in the area between the dashed lines in (a) and (c).}
  \label{temp_2d}
\end{figure}

We conducted 2D PIC simulations to see the multi-dimensional aspect of the radiation cooling effect.
A solid silver foil was placed at the region of $x=21 - 22$\,${\rm \mu m}$ initially.
The initial charge state of the silver ion is 1.
We use the laser parameter of $a_0=2$ and $\tau_l=3$\,ps, which shows the significant radiation cooling during the laser irradiation in the 1D PIC simulation. 
The laser light is focused with a spot diameter $3\,{\rm \mu m}$ normally at the center ($y=10\,{\rm \mu m}$).
The laser electric field oscillates in the $y$-direction, namely, the laser is p-polarized. 

Figure\,\ref{temp_2d} shows the electron temperature distribution with (a), (b) and without (c), (d) radiations.
Figures\,(a) and (c) show the temperature distribution at the moment when the peak of laser arrived at the foil surface. 
In front of the target $x<21\,{\rm \mu m}$, the pre-plasma is heated to $T_e > 10\,{\rm keV}$. 
The electron temperature in (a) and (c) have almost identical distribution in the pre-plasma region. 
Since the emissivity is proportional to the plasma density, e.g., bremsstrahlung depends on the square of the plasma density as discussed in Sec.\,\ref{sec-xray}, the radiation cooling effect is small in the low density region.

Electron temperature inside the foil increases along the incident laser axis via the resistive heating, and reaches $4\,{\rm keV}$ with radiations and $8\,{\rm keV}$ without radiations. 
The heat propagates from the interaction surface via the thermal diffusion.
Overall, as expected from the 1D calculation results, the temperature of the case with radiation is lower than that without radiation.
These results are consistent quantitatively with 1D results as shown in Fig.\,\ref{e_temp}.
(b) and (d) show the temporal evolution of the electron temperature distribution in the $y$ direction. 
Here, the temperature is averaged inside the foil. 
In the case with radiation, the peak temperature was reduced to $4\,{\rm keV}$.
On the other hand, in the case without radiation, the temperature continued increasing during the laser irradiation.
Comparing (a) and (c), we see that the target expansion front slows down with the radiation cooling. 
This is due to that plasma expansion speed is given approximately by the ion acoustic velocity $C_{s}$ which is proportional to $T_{e}^{1/2}$. 
Since the expanding plasma density is low, the radiation cooling is not significant, and consequently, the temperature at the rear side was higher than that inside of the solid density foil.

\begin{figure}[t!]
 \centering
 \includegraphics[width=8cm]{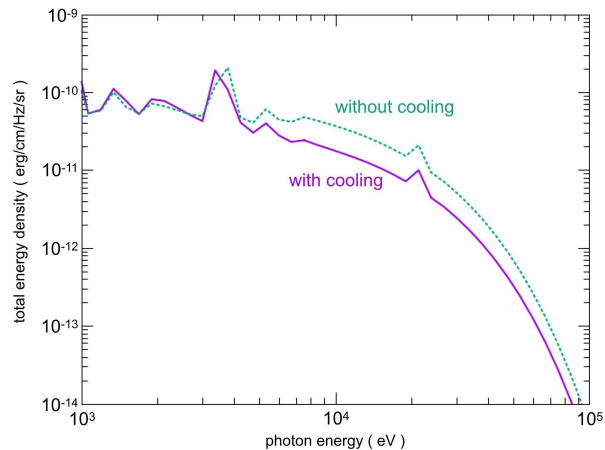}
  \caption{Comparison of the areal density of the radiation energy in the 2D PIC simulations with radiation cooling (solid line) and without radiation cooling (dashed line). The energy was calculated by integrating the emissivity of each photon over the laser spot region inside the foil during the interaction period with the laser. 
}
  \label{rad_spect_2d}
\end{figure}

Figure\,\ref{rad_spect_2d} shows spectra of the areal density of the radiation energy $E_{\rm tot}$ during the laser plasma interaction period $T_{\rm LPI}$ calculated as
\begin{eqnarray}
 {E_{\rm tot}(h\nu_{\rm i}) }\ = \int_{ T_{\rm LPI}} \int\int_{S} \eta\left(x,y,h\nu_{i}\right)\times4\pi \,dxdydt ,
\label{tot_rad_e}
\end{eqnarray}
where $\eta$ is the emissivity.
The spatial integral is taken over the laser spot region inside the foil $S$, i.e., $x=21-22\,{\rm \mu m}$ and $y=8.5-11.5\,{\rm \mu m}$.
The solid line represents the sum of the emissivity from the plasma in the 2D simulation with the radiation cooling.
The dashed line shows the result when the cooling effect is neglected during the interaction.
Below the photon energy of $4\,{\rm keV}$ which mainly comes from plasma at temperature below 1\,{\rm keV}, the two lines are similar. 
Above the photon energy of $4\,{\rm keV}$, the emission energy without radiation cooling is about 2 times higher.
The difference was due to the difference in the temperature distribution seen in Figs.\,\ref{temp_2d} (b) and (d), namely, the higher temperature in (b) results larger hard X-ray emissions.
This means that, without considering the radiation cooling effect, the emission energy will be overestimated especially for higher energy X-rays.

\section{Conclusions}
We studied the dynamics of the ultrafast heating of a silver thin foil by relativistic petawatt laser lights, which have the constant energy, by using the collisional PIC simulations with the radiation cooling. 
The silver plasma heated to a keV-level electron temperature dissipates its energy by hard X-ray emissions in the time scale of a few picoseconds.  
When the laser pulse duration is shorter than the picosecond order and the intensity is over $10^{19}\,{\rm W/cm^{2}}$, the radiation cooling effect on the plasma heating is negligible. 
On the other hand, the radiation cooling affects significantly during the laser irradiation with an over-picosecond pulse and intensity less than $10^{19}$\,W/cm$^2$. 
The radiation cooling decreased the electron temperature and suppressed hard X-ray emission, and therefore, the hard X-ray emission will be overestimated when the radiation cooling effect is neglected.
To consider the radiation cooling is also important in simulating the plasma dynamics. 
Since the bulk plasma can be compressed more easily with the radiation cooling, a highly compressed collisional shock is formed in the dense plasma.
It has been observed that thermal diffusion was suppressed at the compressed region by the strong radiation cooling.

In the current study, the silver plasma is thin enough to be optically thin, so that hard X-ray photons escape without reabsorption. 
If the target becomes thicker, we need to consider the radiation transport. 
We leave it as our future work to see how the radiation transport affects the energy transport in dense plasmas.

\section*{Acknowledgments}
This study was supported by the JSPS KAKENHI grant number JP19KK0072, JP20H00140, JP20K14439, and the Foundation for the Promotion of Ion Engineering.

\end{document}